# Se Nano-Powder Conversion into Lubricious 2D Selenide Layers by Tribochemical Reactions

*Philipp G. Grützmacher[1]\*, Michele Cutini[2], Edoardo Marquis[2], Manel Rodríguez Ripoll[3], Helmut Riedl[4], Philip Kutrowatz[4], Stefan Bug[1], Chia-Jui Hsu[1], Johannes Bernardi[5], Maria Clelia Righi[2]\*, Carsten Gachot[1]\* and Ali Erdemir[6]\**

[1] P. G. Grützmacher, S. Bug, C. J. Hsu, C. Gachot
Institute for Engineering Design and Product Development, Tribology Research Division, TU Wien, 1060 Vienna, Austria
\*E-mail: philipp.gruetzmacher@tuwien.ac.at; carsten.gachot@tuwien.ac.at

[2] M. Cutini, E. Marquis, M. C. Righi
Department of Physics and Astronomy, Alma Mater Studiorum − University of Bologna, Bologna 40127, Italy
\*E-mail: clelia.righi@unibo.it

[3] M. Rodriguez Ripoll
AC2T research GmbH, 2700 Wiener Neustadt, Austria

[4] H. Riedl, P. Kutrowatz
Institute of Materials Science and Technology, TU Wien, 1060 Vienna, Austria

[5] J. Bernardi
University Service Centre for Transmission Electron Microscopy (USTEM), TU Wien, 1040 Vienna, Austria

[6] A. Erdemir
J. Mike Walker '66 Department of Mechanical Engineering, Texas A&M University, College Station, Texas 77843, United States
\*E-Mail: aerdemir@tamu.edu

# Abstract

Transition metal dichalcogenide (TMD) coatings have attracted enormous scientific and industrial interest due to their outstanding tribological behavior. The paradigmatic example is $MoS_2$, even though selenides and tellurides have demonstrated superior tribological properties. Here, we describe an innovative *in-operando* conversion of Se nano-powders into lubricious 2D selenides by sprinkling them onto sliding metallic surfaces coated with Mo and W thin films. Advanced material characterization confirms the tribochemical formation of a thin tribofilm containing selenides, reducing the coefficient of friction down to below 0.1 in ambient air, levels typically reached using fully formulated oils. *Ab initio* molecular dynamics simulations under tribological conditions reveal the atomistic mechanisms that result in shear-induced synthesis of selenide monolayers from nano-powders. The use of Se nano-powder provides thermal stability and prevents outgassing in vacuum environments. Additionally, the high reactivity of the Se nano-powder with the transition metal coating in the conditions prevailing in the contact interface yields highly reproducible results, making it particularly suitable for the replenishment of sliding components with solid lubricants, avoiding the long-lasting problem of TMD-lubricity degradation caused by environmental molecules. The suggested straightforward approach demonstrates an unconventional and smart way to synthesize TMDs *in-operando* and exploit their friction- and wear reducing impact.

# 1. Introduction

The human body with moving joints in relative motion can be considered a highly complex machine. Friction in the human body occurs every time we take a step and even every time we blink. The proper functionality of our body depends on a good lubrication of its moving joints to effectively protect them against damage and wear. Maintaining low friction and wear in the body's tribological contacts involves the biosynthesis and replenishment of biomolecules acting as natural lubricants [1]. In this context, it has been suggested that this biosynthesis can be induced by shear force. Therefore, the relative motion of tribological contacts in the body results in the *in-operando* formation of lubricants, leading to low wear, prevention of joint degradation, and maintenance of tissue function continuously over several decades.

The *in-operando* formation of a lubricating layer protecting the surfaces from wear has many advantages, such as their selective formation at locations where they are needed the most, as well as basically infinite wear life as long as the precursors for their formation are still available. Wear life has a particularly huge impact on the effectiveness of machine components as 70% of their failures can be traced back to surface and interface-specific degradations [2]. In engineering applications, it is quite common to form surface-protecting tribofilms *in-operando* by the decomposition of precursor molecules, such as is happening with the well-known friction modifier molybdenum dithiocarbamate (MoDTC) [3] and the anti-wear agent zinc dialkyldithiophosphate (ZDDP) [4], which has been used as lubricant additives for many decades. A major concern of such additives is that they often rely on organometallic compounds that are noxious for the environment.

Alternatively, it has been shown in the past that it is possible to form solid lubricants based on nanocarbons *in-operando* from hydrocarbon gases or liquids [5,6]. Therefore, mechanochemical processes such as high-energy ball milling can be used [7] or the solid lubricants can be directly formed during tribological loading as a result of high contact pressure, shear, and flash heating [2,8–10]. In this case, the metallic elements required for the *in-operando* formation of the solid lubricants are environmentally compatible as they can be used in metallic form or as inorganic compounds.

Transition metal dichalcogenides (TMDs) (*e.g.*, $MoS_2$, $WS_2$ or $MoSe_2$) have been used by industry for many years, mainly because of their unique tribological properties. Most of them are synthetically produced, but $MoS_2$ exists naturally in a crystalline form, known as molybdenite mineral. Under dry sliding conditions in high vacuum or in inert gas environments they can provide some of the lowest coefficients of friction (COF). Mainly because of such excellent lubrication properties, they are the choice of lubricants for many types of space applications, such as the recently launched James Webb Space Telescope [11].

Their excellent lubricating properties stem from a layered or lamellar structure, in which individual monolayers are composed of transition metal atoms (*e.g.,* Mo and W) sandwiched between the chalcogen atoms (*e.g.,* S and Se). The atoms in each layer are bonded together *via* strong covalent bonds, while the layers themselves are spread relatively wide apart (*e.g.*, $MoS_2$ has an interlayer spacing of 0.615 nm) and are held mostly by weak van der Waals forces. Under shear conditions layers slide over one another to provide low friction while the very strong and resilient layers protect surfaces against wear. The most used TMD-based solid lubricant is $MoS_2$ even though selenides and tellurides have been recently reported to exhibit superior tribological properties [12].

TMDs can be used in the form of solid lubricant coatings, as fillers in the making of self-lubricating polymers, or dispersed in oils and greases to achieve superior friction and wear properties. In most practical applications, TMDs are either burnished to the surfaces to be lubricated or put on as a thin solid film using, for example, sputtering techniques (as in bearings and gear parts of spacecrafts) [13–15]. Powder forms or consolidated compacts with special binders can also be used in certain applications to lower friction. Independent of their form of application, TMDs are transferred to the friction pairs during tribological loading and form a thin layer (very often only a few nm thick) which is mostly responsible for reducing friction and wear [9,16].

The lubricity of most TMDs is strongly affected by humidity, temperature, and other environmental species such as oxygen. In general, higher humidity causes high friction and wear [17–20]. Environmental oxygen may also influence TMDs friction and wear properties, most often triggering oxidation of the metal atoms, giving rise to third body wear due to harder oxides of these metals. In contrast, they work incredibly well in the absence of such environmental species. When operated in high vacuum or space environments, inert gasses like Ar, He, or $N_2$, friction coefficients approaching 0.001 are reported (especially when used in the form of 2D sheets) [21].

In this paper, we describe an innovative and very unique approach that results in the formation of TMDs even with the use of the chalcogen element Se in the form of a nano-powder, sprinkled onto the involved sliding surfaces. The tribochemical reactions in the contact zone result in very low friction and wear. Transmission electron microscopy in combination with Raman and XPS provide some physical evidence for the solid-state formation and the structural and chemical nature of these films.

In the past, the *in-operando* formation of crystalline TMD from amorphous coatings or molecules, such as $H_2S$ and MoDTC, was demonstrated both by experiments [22-27] and simulations [28-30]. However, to the best of our knowledge, no work has previously shown that a powder can be converted in real time into a layered material by sprinkling it into a sliding metal. This idea, which is radically different from previous observations, can open the way to new technological applications due to the advantages for the use in high temperature conditions or at the nanoscale, where liquid media or sputtered coatings are not suitable. This constitutes a significant advancement in the tribology field. Moreover, it can inspire the in-situ synthesis of new compounds even different from TMD.

The solid-solid state formation of such self-lubricating tribofilms seems rather unique as our *ab initio* molecular dynamics simulations (AIMD) show and, therefore, help to understand the underlying mechanisms involved in the *in-operando* formation of such tribofilms. AIMD simulations allows for an accurate description of the formation/dissociation and hybridization-changes in the chemical bonds caused by the mechanical stresses applied. The enhanced reactivity imposed by the tribological conditions is often dictated by quantum mechanical effects that cannot be captured by classical molecular dynamics. The simulations presented here demonstrate that accurate *in silico* experiments are feasible nowadays by exploiting HPC resources, paving the way to a mechanical synthesis of materials with atomistic understanding. The *in-operando* formation of 2D solid lubricants at the contact interface brings many advantages, such as their selective formation at tribological hotspots. Furthermore, the amount of TMDs formed is automatically adapted to the operational conditions, and the use of just the chalcogenide powder instead of the TMD layers prevents the degradation of the lubricant properties by the interaction with air humidity.

# 2. Results and Discussion
## 2.1. Tribological Experiments

We performed unidirectional ball-on-disk experiments using chemically inert $Al_2O_3$ counterbodies to study the *in-operando* formation of solid lubricants based on TMDs, in particular $MoSe_2$ and $WSe_2$. Without the application of Se nano-powders on the sliding surfaces, the evolution of the COF over the sliding cycles is characteristic for non-lubricated sliding of a metal substrate *versus* a ceramic counterbody (**Figure 1**). The COF starts at 0.4 for the bare Mo coating and at 0.3 for bare W and then quickly increases to a maximum value of 0.8 (Mo) and 0.6 (W) during the first 100 sliding cycles of running-in, which can be attributed to the evolution of the real contact area as well as the removal of contaminants and thin native oxide layers, resulting in higher adhesion and finally increased friction [31]. Subsequently, the COF slightly decreases until it stabilizes at a steady-state-value of 0.6 or 0.4 for Mo and W, respectively. The smaller COF of W compared to Mo can be mainly traced back to its higher hardness leading to a smaller real contact area and less disruption of the native oxide film [32]. In contrast, the friction behavior drastically differs when Se nano-powder is added to the tribological interface before the tests. For both Mo and W substrates, the COF starts at a level like that of the reference measurements. However, independently of the used transition metal substrate, the COF continuously decreases over the course of the sliding experiment, ending up at a COF of 0.09 and 0.13 for the Mo and W substrates with Se nano-powder addition, respectively. It is remarkable to achieve such a low macroscopic COF on an unlubricated metal-on-ceramic sliding contact under ambient conditions by simply sprinkling the contact zone with nano-powder.

This improvement in the tribological performance as a result of the addition of Se nano-powder to the contact zone is also reflected in the wear of the substrates (**Figure 1c-f**). When the experiments were performed with Se, the wear rate (as calculated from Eq. 1) decreased by an order of magnitude from $1.1 \cdot 10^{-4}$ to $2.7 \cdot 10^{-5}$ mm$^3$ N$^{-1}$m$^{-1}$ and $5.3 \cdot 10^{-6}$ to $3.9 \cdot 10^{-6}$ mm$^3$ N$^{-1}$m$^{-1}$ for Mo and W, respectively. We observed a clear reduction in wear in terms of wear groove shape and size, wear particle formation, and abrasive wear marks for the experiments performed with Se nano-powder. Again, due to its higher mechanical hardness, W shows much lower wear in general compared to the Mo substrates when rubbing against the $Al_2O_3$ ball. This confirms a less conformal contact with a smaller contact area, thus leading to lower friction.

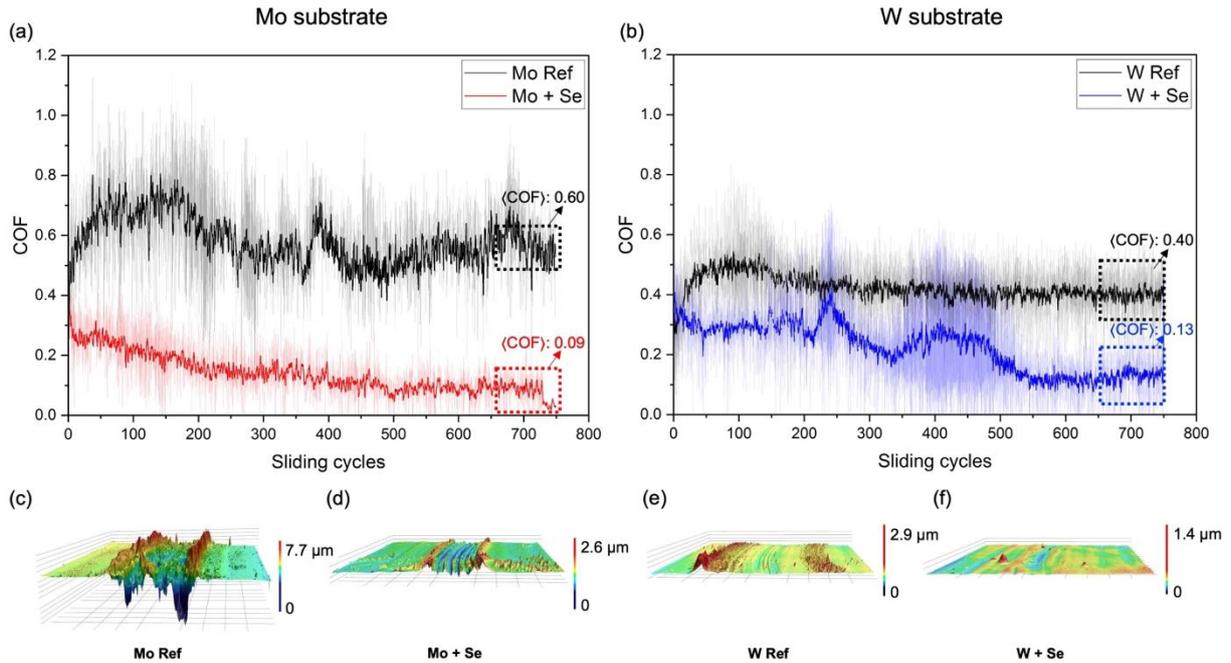

**Figure 1.** (a,b) Friction and (c-f) wear performance of the **Mo** and **W** substrates in ball-on-disk experiments. Compared to the bare substrates (black curves), which show the typical frictional performance of a dry contact, the experiments with Se nano-powder addition (red curves) demonstrate a continuous decrease in COF and a significant reduction in steady-state friction for both (a) **Mo** and (b) **W** substrates. Additionally, the (c,d) **Mo** and (e,f) **W** substrates demonstrate greatly reduced abrasive wear marks when Se nano-powder is added as measured by 3D laser scanning microscopy. Note that all wear tracks were imaged at the same magnification.

To understand the tribochemical reactions taking place in the contact zone during relative sliding, we analyzed the substrate surfaces by Raman spectroscopy, XPS, and TEM. Raman analysis in **Figure 2** unambiguously shows that sliding induced tribochemical reactions occur in the contact zone. As expected, when performing experiments in ambient air atmosphere, sliding on the bare transition metal substrates is dominated by tribo-oxidation of Mo and W, leading to the formation of $MoO_3/MoO_2$ and $WO_3$, which were found in the wear tracks of the Mo and W samples, respectively (**Figure 2b** and **d**). However, many of the oxide peaks disappeared for the cases when Se nano-powder was added to the contact zone of the Mo (**Figure 2b**) and W (**Figure 2d**) samples. Instead, peaks demonstrating the formation of $MoSe_2$ and $WSe_2$ became very clear. This statement is not entirely conclusive, as the peaks of $MoO_2$ at 210 cm$^{-1}$ and $MoSe_2$ at 242 cm$^{-1}$ as well as $WO_3$ at 265 cm$^{-1}$ and $WSe_2$ at 245 cm$^{-1}$ partly overlap. Nevertheless, these findings point towards a tribochemical reaction between the transition metal substrates and the Se nano-powder induced by the high mechanical stresses and shear forces during sliding, by which the lubricating TMDs $MoSe_2$ and $WSe_2$ are formed. The formation of a tribofilm on the substrates' surfaces consisting of these 2D layered TMDs results in the reduction of friction and wear due to their easy-to-shear ability and the protection of the underlying surface.

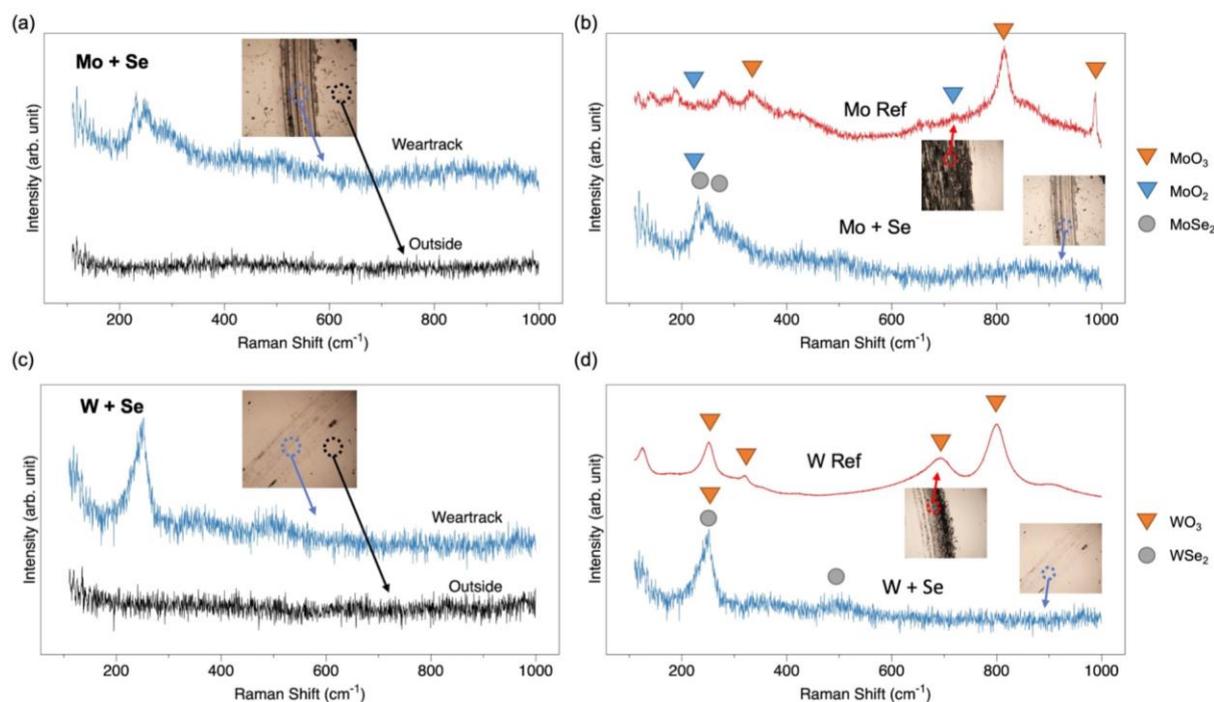

**Figure 2.** Raman analysis of the wear tracks of the (a,b) Mo and (c,d) W substrates demonstrates the tribochemical reactions occurring in the contact area induced by the relative motion between substrate and counterbody. The addition of Se nano-powder triggers chemical reactions on both (a) Mo and (c) W substrates. On both substrates (b,d) selenides are formed during sliding, whereas without Se nano-powder addition oxide formation dominates.

A more detailed investigation of the topmost surfaces' chemical composition was performed by XPS. The high-resolution XPS scans performed inside the wear tracks are presented in **Figure 3**. The wear tracks from experiments performed in standard atmosphere using Mo substrates reveal the complex XPS envelope characteristic of molybdenum oxide samples containing mixed oxidation states. The main peak, located at 228.0 eV, is attributed to metallic Mo. The Mo $3d_{5/2}$ peaks are located at 230.8, and 232.6 eV and are assigned to the oxidation states Mo (IV) and Mo (VI) [33]. The presence of Mo (V) corresponding to $Mo_2O_5$, or more in general to non-stochiometric Magnéli phases of the form $Mo_nO_{3n-1}$, could only be measured after sputtering for 10 s, once the intensity of the Mo (VI) peak was drastically reduced (Supporting Information, **Figure S1**).

The addition of Se powder to the contact area before the experiment results in a radical change in the chemical composition of the wear track. Aside from the major presence of metallic Mo (228.0 eV), the spectrum of the as-tested wear track reveals a doublet of Mo $3d_{5/2}$ and Mo $3d_{3/2}$ centered at 228.8 and 231.8 eV, respectively. These peaks are characteristic of $MoSe_2$, [34].

Additionally, the presence of Mo (IV) and Mo (VI) can be identified. The spectrum obtained after 10 s sputtering is analogous and shows a very similar composition, with the main difference that the presence of $MoO_3$ is almost completely removed (Supporting Information, **Figure S1**). Outside the wear scar, the surface composition was determined to be formed by metallic Mo and $MoO_3$, with a minor presence of $MoO_2$. $MoO_3$ was completely removed after 10 s sputtering. The *in-operando* formation of $MoSe_2$ during the experiment is further confirmed by the presence of the characteristic selenide doublet at 54.6 eV (Se $3d_{5/2}$) and 55.4 eV (Se $3d_{3/2}$) [35,36]. The XPS measurements performed on the W substrates offer analogous results. The XPS spectrum measured on the W substrate tested without Se shows a main doublet at 31.2 and 33.4 that corresponds to W (0). The other peaks are ascribed to tungsten oxides ($WO_2$ and $WO_3$) [37]. In the presence of Se nano-powder, the wear track shows a similar spectrum but with a pronounced doublet at 32.4 and 34.2 that corresponds to $WSe_2$. After sputtering the sample for 10 s, the peaks corresponding to $WSe_2$ and $WO_3$ dramatically decline, indicating that these compounds are found at the outermost surface (Supporting Information, **Figure S1**). Outside the wear scar, the measured spectrum contains only peaks corresponding to W (0), W (IV) and W (VI). The Se narrow spectra again confirm the presence of $WSe_2$ with a main peak at 54.7 eV, accompanied in this case by the presence of a metallic tungsten satellite peak located at 56 eV [38]. In this case, the Se envelope is less pronounced when compared to the scan performed on the Mo substrate. After sputtering for 10 s, the selenide doublet significantly diminishes in agreement with the measurements described for the Mo narrow scan.

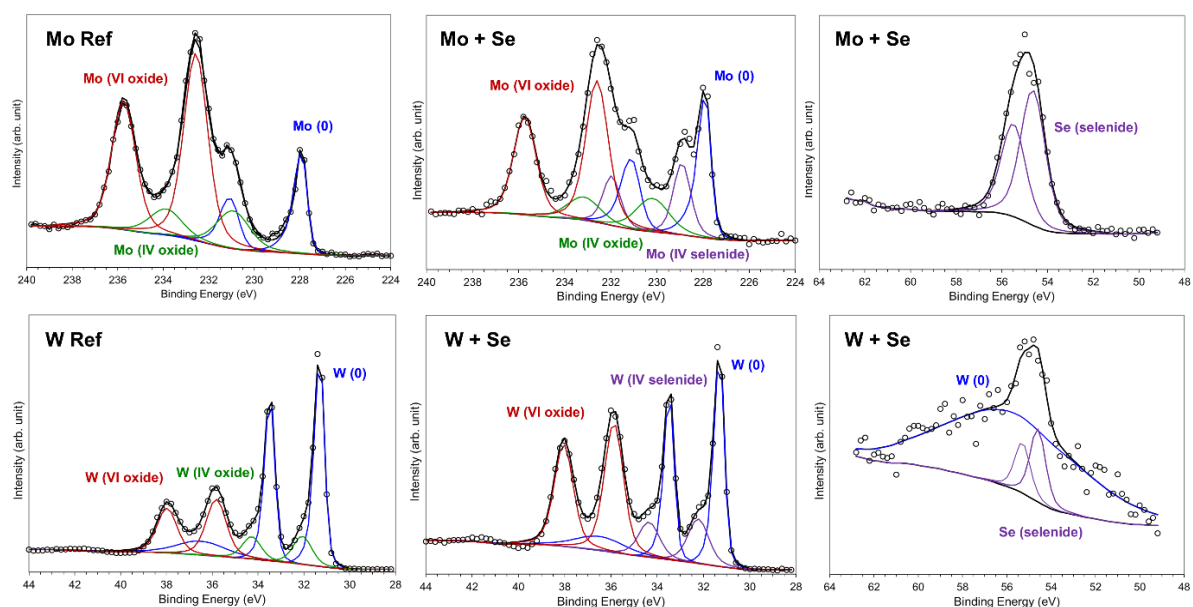

**Figure 3.** High-resolution XPS spectra of Mo 3d, W 3d and Se 2p illustrating the presence of $MoSe_2$ (Mo $3d_{5/2}$ and Mo $3d_{3/2}$ centered at 228.8 and 231.8 eV) and $WSe_2$ (doublet at 32.4 and 34.2 eV). The formation of selenides is confirmed by the presence of a doublet at 54.6 eV (Se $3d_{5/2}$) and 55.4 eV (Se $3d_{3/2}$). Note the lack of selenide formation in the reference samples.

Finally, the structure of the formed tribofilms on the Mo and W substrates were investigated by TEM. The results in **Figure 4** clearly verify the formation of 2D layered transition metal selenides on both substrates. The tribofilm found on the Mo substrate has a thickness of up to 20 nm and is composed of clearly layered structures intermixed with amorphous zones. The tribofilm on the W substrate indicates a thickness of up to 10 nm and is, therefore, much thinner compared to the one on the Mo substrate. However, the tribofilm is more clearly delineated from the substrate and is composed of continuous layers. This can be explained by the larger resistance to plastic deformation of W, which prevents strong intermixing as found for Mo. The measured interlayer distances of ~7.1 Å is close to the ones corresponding to the (0 0 2) planes of hexagonal $MoSe_2$ and $WSe_2$ [39,40]. Additionally, these measurements are backed up by EDS measurements at the TEM lamellae, which clearly demonstrate an accumulation of Se at the topmost surface of the wear track (Supporting Information, **Figure S2** and **Figure S3**). In the case of Mo, elemental mappings by TEM-EDS furthermore demonstrate that Se is continuously present across the wear track's surface, thus verifying the continuity of the tribofilm based on $MoSe_2$ (Supporting Information, **Figure S4**).

The finding of layered structures in the tribofilm combined with the Raman, TEM-EDS, and XPS results clearly proves the *in-operando* formation of low friction and low wear tribofilms based on transition metal selenides. Additionally, the layers are perfectly aligned with respect to the substrate surface, thus enabling easy shearing and a low COF.

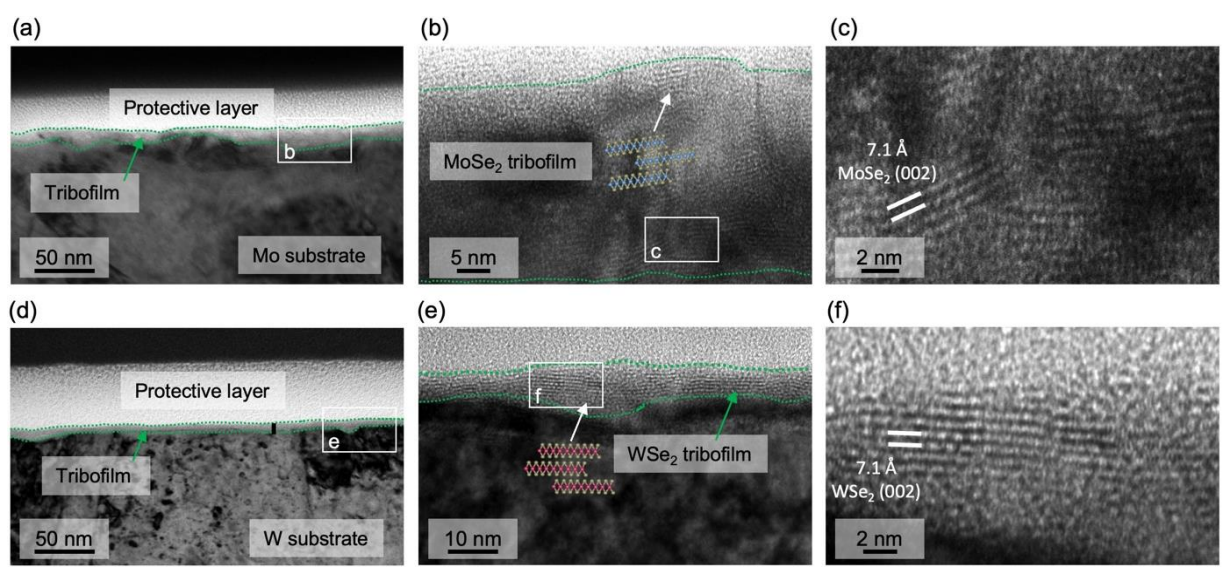

**Figure 4.** TEM micrographs of the tribofilms formed during the tribological experiments on the (a-c) Mo and (d-f) W substrates with addition of Se nano-powder. The tribofilms are encircled with dashed green lines and the layer distances are measured and given in the high magnification images.

The re-establishment of a low coefficient of friction was studied using Mo and W coatings where the Se nano-powder was sprinkled again into the tribological contact zone after the COF had reached the high value observed at the beginning of the tests. This increase in friction can be correlated to a degradation of the initially formed tribofilm.

In accordance to the measurements discussed before, the COF of the Mo-coated substrate drops after the first addition of Se powder and keeps constant up to around 700 sliding cycles (**Figure 5a**). Thereafter, frictional instabilities appear and the COF increases to 0.35 at approximately 1700 sliding cycles. This is exactly the point when the tribometer was stopped and the Se nano-powder was added again to the contact interface. It can be clearly observed that the COF rapidly decreases and reaches a rather constant level of about 0.1 up to 6500 sliding cycles, which also denotes the end of the test. In **Figure 5b**, the same trend is visible for the W-coated substrate. After the initial application of Se powder, the COF is around 0.17 up to 700 sliding cycles but then starts to increase to even 0.5 at around 850 cycles. After adding the Se nano-powder for the second time, the COF sharply drops down to approximately 0.17 again, where it remains constant up to 1200 cycles before slowly rising again. This straightforward experiment clearly shows the consistency of the approach, demonstrating that a further addition of Se nano-powder results in a repeatable and reproducible friction reduction. It can be envisioned that this cycle could be repeated again various times.

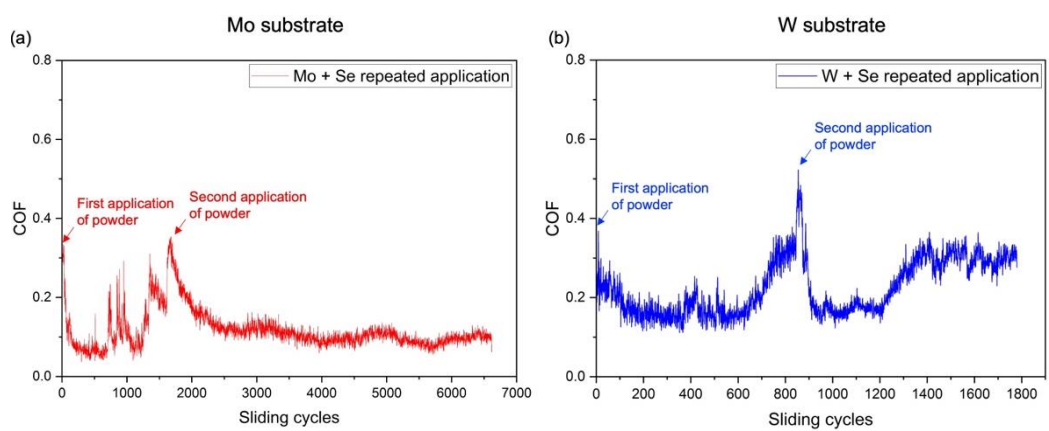

**Figure 5.** Reapplication of Se nano-powder leads to reoccurring reduction in friction.

## 2.2. Mechanism of TMDs Formation

As a first step to understanding the atomistic mechanism of TMD mechano-synthesis, we evaluated if the formation of TMDs from their metal and chalcogenides constituents is energetically favorable. To this aim, we compared the energy of a $MoSe_2$ ($WSe_2$) unit in the TMD layer with the sum of the energies of a Mo (W) atom and two Se atoms in their bulk structures. The results indicate that forming a TMD layer from the bulk constituents is highly favorable, the associated energy gain being 2.0 eV (1.1 eV) per $MoSe_2$ ($WSe_2$) unit.

The energetic analysis described above points at the thermodynamic driving force for the transformation of the metal and chalcogenide elements into the TMD layers. To "observe" such a transformation in real time we performed AIMD simulations of Se nano-powder under load and shear in the presence of a metal debris particle. Since the material structure may be weakened by oxidation, as revealed by Raman analysis (**Figure 2**), this might favor the formation of debris particles. The load and shear were applied through two TMD boundary layers to reduce the computational effort and to operate with the proper in-plane cell size to host a TMD layer that may be formed. The ball-and-stick representation of these system is reported in **Figure S6** in the Supporting Information, while **Figure 6a** shows different snapshots acquired during the AIMD simulation of sliding.

The initial configuration (t = 0 ps) corresponds to the optimized system structure under load. The high applied load induces the weak Se-Se bonds to rearrange and adhere to the metallic debris particles and the boundary TMD surfaces. We then observe a sliding-induced atomic intermixing. Mo-Se bonds are established with trigonal prismatic coordination typical of TMDs. Already after 3.5 ps, one Mo atom (highlighted with a black circle) is fully surrounded by six Se atoms, indicating a complete detachment from the metallic debris particle. After 35 ps of simulation, the metallic particle is almost completely disaggregated, and only a few metal-metal bonds are still present. At 136 ps the formation of the TMD is almost complete.

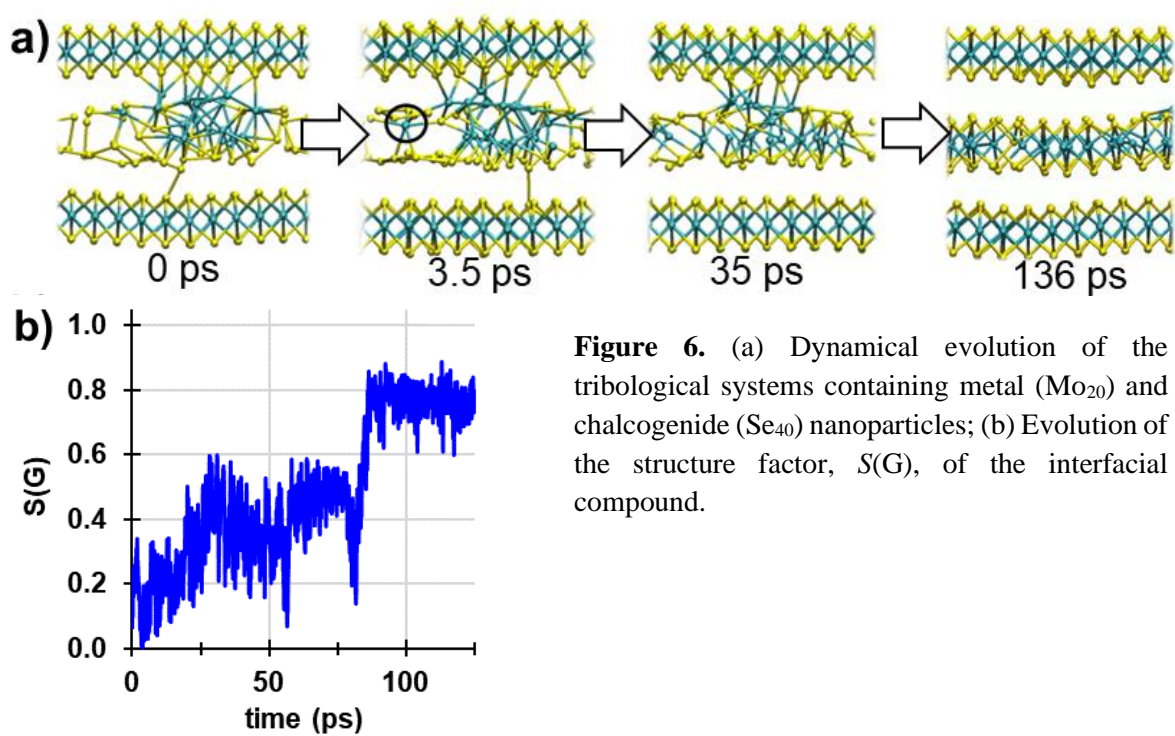

**Figure 6.** (a) Dynamical evolution of the tribological systems containing metal ($Mo_{20}$) and chalcogenide ($Se_{40}$) nanoparticles; (b) Evolution of the structure factor, $S(G)$, of the interfacial compound.

The formation of TMD can be monitored quantitatively by computing the structure factor S(G) of the Mo atoms:

$$S(G) = \sqrt{\frac{\left|\sum_{j=1}^{N} e^{iGr_j}\right|^2}{N^2}}, \qquad (1)$$

where G is the summation of the two reciprocal lattice basis vectors of a crystalline $MoSe_2$ layer, and $r_j$ is the real space position of the j-th Mo atom belonging to the debris particle or the Mo(110) surface. S(G) has a limit value of 0 in case of full incommensurability, and of 1 for full commensurability of the lattices [41]. In **Figure 6b**, we report the S(G) evolution with time for the AIMD performed in this work. The structure factor increases during the simulation, indicating that the atoms belonging to the metal particle reorganize into an almost perfect $MoSe_2$ crystal structure within 80 ps (blue line).

The above-described simulations indicate that mechanical load and especially shearing can promote the formation of crystalline $MoSe_2$ layers from a Se nano-powder as a precursor. This process occurs more easily when metal particles are extracted from the substrate. Wear during running-in can induce this extraction of metal particles, which might be further accelerated by defects or (oxidic) debris particles produced during the sliding process. When metal wear debris particles are formed and brought into contact with the Se nanoparticles, TMD layers can form rapidly under shear and load. The proposed lubrication mechanism based on the *in-operando* formation of TMD layers is schematically shown in **Figure 7**.

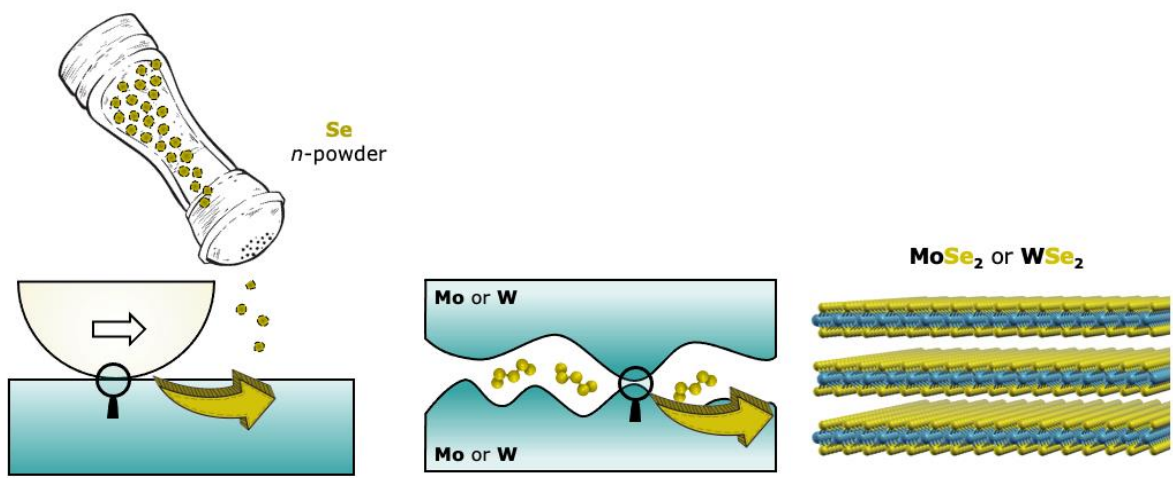

**Figure 7.** Schematic representation of the proposed lubrication method based on the friction-induced formation of slippery TMD layers *in-operando*.

# 3. Conclusion

In this research work, we presented an innovative lubrication method to synthesize slippery TMD layers *in-operando* by simply sprinkling a nano-powder of chalcogen elements, here selenium, at the tribological interface in ambient air atmosphere. The formation of these TMD layers in the tribological contact results in low friction and wear, reducing the coefficient of friction down to 0.1 or even below, levels typically reached only with fully formulated oils.

Without applying the nano-powder, both friction and wear are high, as expected for a dry contact between metal and ceramic, due to abrasive and adhesive material interactions between the sliding surfaces caused by harder oxides of Mo and W.

High-resolution characterization techniques such as Raman- and X-ray photoelectron spectroscopy clearly confirmed the formation of $MoSe_2$ and $WSe_2$ tribofilms on the interacting surfaces. Transmission electron microscopy even revealed the layered nanoscopic structure of $MoSe_2$ and $WSe_2$ with a preferential crystalline alignment with respect to the sliding surfaces.

Fundamental insight into the formation of the selenide monolayers is gained using *ab initio* molecular dynamic simulations, which reveal the shear driven reaction pathways involved in the *in-operando* formation of $MoSe_2$ and $WSe_2$ during sliding. The simulations clearly emphasize that metallic debris are a prerequisite to facilitate the tribochemical reaction between the metallic particles and the Se nano-powder, resulting in the formation of crystalline TMD layers. The metallic debris particles are most likely formed during the sliding of surfaces, which are weakened by oxidation.

Overall, our study demonstrates that the extremely favorable *in-operando* formation of highly lubricous transition metal dichalcogenides through tribochemical reactions between Mo/W and solid Se nano-powder is feasible. This straightforward and cost-effective approach leads to highly reproducible results that reflect the high affinity of the Se nano-powder for tribochemical formation of 2D selenides in the presence of a transition metal coating once they are brought together in the contact interface. Our approach may be considered for applications where poisonous gaseous reactants like $H_2S$ are neither feasible nor desirable, in vacuum environments, where outgassing of conventional oils and greases may prevent their application or in maintenance-critical applications, where commonly used TMD coatings lead to component failure once they are fully worn. The results presented here may inspire more research into solid-state tribochemistry and the design of new composite materials that can provide long-lasting, self-sustained, and self-healing capabilities in future applications *via* the replenishment of a sliding contact with solid lubricants.

# 4. Methods

*Materials and Coating Deposition*

As substrate materials polished AISI 52100 steel platelets (20 x 10 x 1 mm³) were used. Before the coating process, all samples were ultrasonically pre-cleaned in acetone and isopropanol, respectively. Metallic W and Mo coatings with a thickness of around 2 to 3 µm were synthesized from respective metallic targets (3-inch geometry, Plansee Composite Materials GmbH, 99.3% purity) in pure Ar atmosphere (99.999 % purity) using an in-house developed direct current magnetron sputtering system (base pressure below $2.0 \times 10^{-4}$ Pa). The rotating substrate holder (0.25 Hz) was positioned at a target-to-substrate distance of 110 mm. Following a heating sequence to a substrate temperature of 400 °C, an Ar-ion etching step was performed to finally clean the substrates before deposition – pure Ar atmosphere at 5 Pa and an applied bias potential of -800 V. The thin films were then grown at a total Ar pressure of 0.4 Pa, a target current of 0.4 A, and floating potential. The resulting surface roughness ($S_q$) of the Mo and W coatings was 50 and 46.5 nm, respectively. The hardness of the coatings was 12.6 GPa (W) and 7.1 GPa (Mo) as determined by nanoindentation at 4 mN load from 40 idents.

*Tribological Experiments*

Tribological testing was performed on a ball-on-disk setup (Nanovea T50 Tribometer) in unidirectional sliding mode with a sliding radius of 2 mm. The counterbody was an $Al_2O_3$ sphere with a diameter of 6 mm (surface roughness $S_q$: 40 nm), to avoid substantial wear of the counterbody as well as contamination with other transition metals and catalytic effects of metallic particles. The load was set to 1 N, leading to a peak Hertzian contact pressure of 0.88 GPa and 0.92 GPa for the Mo and W substrates, respectively. The sliding speed was set to 1000 mm/min. The tests were run under ambient air with a temperature of 25 °C and a relative humidity of ~45%. After the sample was fixed in the tribometer, 10 mg of high-purity (99.9 %) Se nano-powder with a particle size distribution between 40 to 80 nm (NANOSHELL UK Ltd.) was manually sprinkled onto the substrate surface with a spatula directly underneath the $Al_2O_3$ counterbody. Subsequently, the counterbody was brought into contact with the Se nano-powder so that there was a visible transfer of the loose Se nano-powder from the substrate to the counterbody. Afterwards, the sliding experiments were started. Additional reference measurements are performed without the addition of Se powder. All tests were run three times and the mean values and standard deviations are plotted. Additional tests with Se nano-powder were performed to study the ability to re-establish lubricious TMD layers after their degradation.

Therefore, the tribometer was stopped once the COF reached the high friction value observed at the beginning of the test, Se nano-powder was applied again to the contact zone analogously to the test before, and the test was restarted again without changing the relative position of substrate and counterbody.

*Characterization of the Tribofilm*

The substrate surfaces and wear tracks were imaged by confocal laser scanning microscopy (CLSM, Keyence VK-X1100) and scanning electron microscopy (SEM). The average cross-sectional area of the wear tracks *A* as measured by CLSM was used to calculate the wear rate using the expression:

$$K \left[\frac{\text{mm}^3}{\text{N} \cdot \text{m}}\right] = \frac{A \cdot L}{F_n \cdot N \cdot L}, \qquad (2)$$

where *L* is the wear track length, $F_n$ is the normal force, and *N* is the number of sliding cycles. The chemical state of the surfaces before and after wear was investigated by Raman spectroscopy (HR800, Jobin Yvon Horiba) with a laser wavelength of 532 nm and a yielded laser power of 13.5 mW was used. The spot size was 20 µm and the measuring time 3 s. Additionally, X-ray photoelectron spectroscopy (XPS, Thermo Fisher Scientific Theta Probe) was used to determine the chemical bonding states in the top-most surface layers. The XPS device was equipped with a monochromatic Al Kα X-ray source (hν = 1486.6 eV) and an $Ar^+$ ion gun. The analyzed spot size diameter was 100 µm, with pass energies of 200 eV for the survey spectra and 50 eV for the high-resolution spectra. The XPS spectra were acquired before and after 10 seconds of fine sputtering which was conducted in order to remove any possible surface contamination. The measurements were performed at a base pressure of $2 \times 10^{-9}$ mbar. All the high-resolution spectra were referenced to the adventitious carbon (C 1s binding energy of 284.8 eV). Afterwards, the spectra were processed with the Thermo Fisher Scientific Avantage Data System using Gaussian-Lorentzian peak fitting. Transmission electron microscopy (TEM, FEI TECNAI F20) was used to image the structure of the formed tribofilms. Energy dispersive X-ray spectroscopy (EDS) inside the TEM was performed using an EDAX detector.

*Ab initio Calculations*

Spin-polarized DFT calculations were performed employing version 6.7 of the Quantum ESPRESSO package [42]. The generalized gradient approximation (GGA) within the Perdew-Burke-Ernzerhof (PBE) parametrization was used to describe the electronic exchange and correlation [43]. Dispersion interactions were included using the D2 parametrization scheme of Grimme [44]. The plane-wave expansion of the electronic wave functions (charge density) was truncated with a cutoff of 40 Ry (320 Ry). The ionic species were described by ultrasoft pseudopotentials, considering 14 explicit electrons for the d-ions (*e.g.*, W and Mo), and 6 explicit electrons for chalcogens (*e.g.*, O and Se). For the structural optimization, we adopted default criteria for energy and forces convergence and used a Gaussian smearing of 0.02 Ry. We modelled the (110) surfaces of Mo and W by sampling the Brillouin zone with a 14x18x1 Monkhorst-Pack grid, while an equivalent sampling was used for larger cells [45]. To avoid spurious interactions, all surfaces and interfaces were built with at least 15 Å of vacuum between vertical replicas.

For the AIMD simulations of sliding surfaces we used a version of the Born-Oppenheimer Molecular Dynamics modified by our group to impose tribological conditions. The outmost metallic atoms are dragged with a constant velocity of 2 Å/ps; a normal load of 10 GPa, and a constant temperature of 600 K are applied. Newton's equations of motion are integrated using the Verlet algorithm. We tuned the time step, $\Delta t$, according to the frequency of bond vibrations, with the Mo-O bond stretching having the shortest period of 33 fs (corresponding to a frequency $\nu = 1000$ cm$^{-1}$) [46]. Therefore, we chose a $\Delta t$ value of 2.4 fs, sampling more than 10 times each bond vibration. This choice allows for a fast production while ensuring a correct integration of the equations of motion. To observe the full formation of TMD from Mo and Se particles, we performed long simulation runs (up to hundreds of ps). In this case, where Mo-O bonds were not present, the $\Delta t$ was increased to 9.4 fs. This value is suitable considering that the period of Mo-Se bond stretching is 95 fs. We validated the accuracy of the results by overlapping the calculated trajectory with the one obtained with $\Delta t=2.4$ for a time period of about 20 ps. The results are reported in **Figure S6** of the Supporting Information.

# Acknowledgements

Part of this work was funded by the Austrian COMET-Program (Project K2 InTribology, Grant No. 872176). The authors are indebted to Dr. C. Tomastik for performing the XPS measurements. A.E. acknowledges the Texas A&M Engineering Experiment Station startup funds and the Governor's University Research Initiative. These results are part of the "Advancing Solid Interface and Lubricants by First Principles Material Design (SLIDE)" project that has received funding from the European Research Council (ERC) under the European Union's Horizon 2020 research and innovation program (Grant agreement No. 865633).

## Author contributions

P.G.G. contributed conceptualization, methodology, investigation, analysis, visualization, writing – original draft; M.C. contributed investigation, methodology, software, visualization; E.M. contributed investigation, methodology, software, visualization; M.R.R. contributed investigation, analysis, visualization, writing – original draft; H.R. contributed investigation, resources, writing – review & editing; P.K. contributed investigation; J.B. contributed investigation; C.G. contributed conceptualization, investigation, resources, supervision, writing – review & editing; M.C.R. contributed conceptualization, investigation, resources, supervision, writing – original draft; A.E. contributed conceptualization, investigation, supervision, writing – original draft.

## Conflict of Interest

The authors declare no financial or non-financial conflicts of interests.

## Data Availability

The data that support the findings of this study are available from the corresponding authors upon request.